\documentclass[12pt,a4paper,twoside]{article}
\usepackage[utf8]{inputenc}
\usepackage[
	left=2.5cm,right=2.5cm,
	top=3cm,bottom=2.5cm,
	]{geometry}
\usepackage{graphicx}
\usepackage[T1]{fontenc}
\usepackage{csquotes} 
\usepackage[ngerman,english]{babel}
\usepackage{hyperref}
\usepackage{amsmath} 
\usepackage{amssymb} 
\usepackage{amsfonts}
\usepackage{array}
\usepackage{bm} 
\usepackage{multirow} 
\usepackage{cite}

\makeatletter
\newcommand\blfootnote[1]{%
  \begingroup
  \renewcommand\thefootnote{}\footnote{#1}%
  \addtocounter{footnote}{-1}%
  \endgroup
}
\makeatother

\usepackage{xspace}
\newcommand{\codename}{\textsc{BSFfast}\xspace}

\newcommand{\BSF}{\text{BSF}}
\newcommand{\vel}{v}

\newcommand{\ab}{\alpha_b}

\newcommand{\E}{\mathrm{e}}
\newcommand{\I}{\mathrm{i}}
\newcommand{\eq}{\text{Eq.}}

\newcommand{\fig}{\text{Fig.}}
\newcommand{\tab}{\text{Tab.}}

\newcommand{\bra}[1]{\langle #1 \vert}
\newcommand{\ket}[1]{\vert #1 \rangle}

\def\be{\begin{equation}}
\def\ee{\end{equation}}
\def\bea{\begin{eqnarray}}
\def\eea{\end{eqnarray}}
\newcommand{\lr}{\left(}
\newcommand{\rr}{\right)}

\newcommand{\el}{\ell}
\newcommand{\lp}{{\el^{\prime}}}
\newcommand{\gbb}{\gamma_n}
\newcommand{\zbb}{\zeta_n}
\newcommand{\zs}{\zeta_s}
\newcommand{\svta}[1]{\left< \sigma_{#1} \vel \right>}

\hypersetup{
	colorlinks=true,
	linkcolor=black, 
	citecolor=black,
	urlcolor=black, 
	pdftitle={Non-perturbative effects in Production and Detection Processes of Dark Matter},
	pdfauthor={Stefan Lederer},
}

\begin{document}

\begin{titlepage}
\begin{flushright}
{\small
TTK-25-50,
TUM-HEP-1589/26
}
\end{flushright}

\vskip1cm


\begin{center}
{\Large\bf
\codename:~Rapid computation of bound-state effects\\
\vspace{0.5ex}
on annihilation in the early Universe}
\end{center}

\vspace{0.1cm}
\begin{center}
{\sc Tobias Binder$^a$, Mathias Garny$^a$, Jan Heisig$^b$, Stefan Lederer$^c$
}\\[0.6cm]
$^a${\it\small Physik Department, 
James-Franck-Stra\ss{}e~1, 
Technische Universit\"at M\"unchen,\\
D--85748 Garching, Germany}\\[0.2cm]
$^b${\it\small Institute for Theoretical Particle Physics and Cosmology, RWTH Aachen University,
Sommerfeldstra\ss e 16,
D-52056 Aachen, Germany}\\[0.2cm]
$^c${\it\small 
National Centre for Nuclear Research, Pasteura 7, 02-093 Warsaw, Poland
}

\end{center}

\vspace{0.6cm}
\begin{abstract}
Bound-state formation (BSF) can have a large impact on annihilation of new physics particles with long-range interactions in the early Universe. In particular, the inclusion of excited bound states has been found to strongly reduce the dark matter abundance and qualitatively modify the associated freeze-out dynamics. While these effects can be captured by an effective annihilation cross section, its explicit computation is numerically expensive and therefore impractical for repeated use in Boltzmann solvers or parameter scans. In this work we present \textsc{BSFfast}, a lightweight numerical tool that provides precomputed, tabulated effective BSF cross sections for a wide class of phenomenologically relevant models, including highly excited bound states and, where applicable, the full network of radiative bound-to-bound transitions. We exploit rescaling relations of the cross section to efficiently cover models with additional free parameters and provide fast interpolation routines in Mathematica, Python and C for use in Boltzmann solvers.
As an illustration, we apply \textsc{BSFfast} to a superWIMP scenario with a colored mediator, demonstrating that the tool enables phenomenological studies that would otherwise be computationally prohibitive. The code is publicly available on \href{https://github.com/bsffast/BSFfast}{GitHub}.$^\star$ 
\end{abstract}

\blfootnote{\footnotesize $^\star$ \href{https://github.com/bsffast/BSFfast}{\url{https://github.com/bsffast/BSFfast}}}

\end{titlepage}

\tableofcontents

\section{Introduction}
\label{sec:intro}

Exploring cosmological consistency is one of the pillars of studying theories beyond the Standard Model, often requiring the computation of particle densities of new species in the early universe. For particles sharing a long-range force, the formation of metastable bound states can significantly affect the dynamics through an effective enhancement of the species’ annihilation rate. In particular, its impact on dark matter genesis has been widely recognised~\cite{Ellis:2015vaa,Ellis:2015vna,Liew:2016hqo,Kim:2016zyy,Harz:2017dlj,Mitridate:2017izz,Harz:2018csl,Biondini:2018pwp,Biondini:2018ovz,Fukuda:2018ufg,Biondini:2019int,Bottaro:2021snn,  Garny:2021qsr,Becker:2022iso,Binder:2023ckj}, be it via dark matter itself forming bound states or via an accompanying particle such as a coannihilator, or a heavy mother particle that freezes out and whose decay produces dark matter out of equilibrium, i.e.~the superWIMP scenario.

More recently, the role of excited states has gained attention. While less important for scenarios of rapid dark matter decoupling, their impact increases drastically the longer the freeze-out process persists. Such a situation arises, for instance, in conversion-driven freeze-out and superWIMP production involving long-lived coloured mediators as well as in thermal freeze-out scenarios in dark sectors where dark matter is charged under a long-range force, in particular a non-Abelian gauge interaction. In such cases, annihilation can remain efficient over a prolonged period and can even render highly excited bound states to become the dominant contribution to the effective annihilation cross section.

As shown in Refs.~\cite{Binder:2021vfo,Garny:2021qsr} (see also \cite{Ellis:2015vna,Mitridate:2017izz} for earlier work based on the same ansatz), the effect of arbitrarily highly excited states can be captured by an effective annihilation cross section of their constituents, $\langle\sigma v \rangle_\text{eff,BSF}$, making their inclusion in the Boltzmann equations straightforward. However, the computation of $\langle\sigma v \rangle_\text{eff,BSF}$ is computationally demanding. The evaluation of a large number of complex scattering-to-bound and bound-to-bound transition rates, together with the inversion of the associated transition network matrices, is very time-consuming. Consequently, such “on-the-fly” computations are far too expensive to be used inside Boltzmann solvers, parameter scans, or phenomenological studies that require repeated evaluation of the effective thermally averaged cross section~$\langle \sigma v\rangle_{\rm eff,\BSF}(x)$.

The purpose of this work is to close this gap. We present \codename, a lightweight numerical tool that provides precomputed, tabulated effective BSF annihilation cross sections for a wide range of masses and model choices. Expanding upon previous works \cite{Garny:2021qsr,Binder:2023ckj, Beneke:2024nxh}, we focus on BSF effects within unbroken gauge theories described by dipole transitions.  
The models covered by \codename include particles charged under SM QCD and/or QED, such as (colour-)charged mediators appearing, for instance, in $t$-channel dark matter models or supersymmetric scenarios, as well as particles charged under a dark QED and dark QCD group with general gauge coupling strength, in all cases for both scalar and fermionic constituents (see Table~\ref{tab:model-coverage} in Sec.~\ref{sec:modelcov} for the full list).

For each case, \codename provides fast interpolation in the temperature parameter $x=m/T$ and in the particle mass, and -- where applicable -- employs powerful rescaling properties that allow one to efficiently span a wide range of gauge couplings without additional tabulation, while consistently accounting for the contribution of highly excited bound states wherever required for convergence. This makes the tool readily usable for integration into  Boltzmann solvers such as \textsc{micrOMEGAs}~\cite{Alguero:2023zol} and \textsc{MadDM}~\cite{Ambrogi:2018jqj}, without the need to rerun the expensive underlying integral computations. 
Very recently, a related approach to including Sommerfeld enhancement and BSF effects in relic-density computations has been presented in Ref.~\cite{Becker:2025vgq}, where these effects are implemented directly within \textsc{micrOMEGAs} for coloured dark sectors. While that work focuses on an on-the-fly evaluation inside the solver framework, \codename adopts a complementary interpolation- and rescaling-based strategy, allowing the inclusion of highly excited bound states up to principle quantum numbers of $n=100$, far beyond the scope of \cite{Becker:2025vgq}.

The remainder of the paper is structured as follows. After reviewing the relevant expressions entering the effective cross section in Sec.~\ref{sec:BSFXS}, we describe important rescaling relations that allow an efficient generalization in Sec.~\ref{sec:rescale}. The tool's coverage of new physics models, details of the numerical implementation, and aspects of partial wave unitarity are discussed in Sec.~\ref{sec:impl}. Finally, in Sec.~\ref{sec:application} we apply \codename to a phenomenological study of the superWIMP scenario, before concluding in Sec.~\ref{sec:concl}. Details on the public implementation and usage are provided in Appendix~\ref{app:usage}.

\section{Effective description of bound-state effects}
\label{sec:BSFXS}

We consider a non-relativistic particle species $X$ with mass $m $ subject to long-range interactions described by either an Abelian $\text{U}(1)$ or non-Abelian $\text{SU}(N)$ unbroken gauge interaction. Here $X$ can stand either for a particle species that comprises (part of) the dark matter, or for a mediator particle whose abundance impacts the dark matter density via e.g. coannihilation, conversion-driven freeze-out or thermal freeze-out followed by decay (i.e.~the superWIMP mechanism), or for a particle species totally unrelated to dark matter, but relevant for other early-Universe processes. If $X$ stands for a mediator state, the gauge interaction can be given by the SM gauge groups\footnote{For the case of QCD, we focus on temperature scales well above the confinement scale, and treated it as a weakly coupled theory with running gauge coupling. We neglect the running of couplings in all other cases.}, in which case the gauge couplings are known. We also consider the possibility of a dark sector, in which case the $\text{U}(1)$ or $\text{SU}(N)$ gauge interaction corresponds to a new dark force beyond the SM, with gauge coupling strength $\alpha\equiv g^2/(4\pi)$ treated as a free model parameter. 

 In Section~\ref{sec:effective}, we briefly review the effective treatment of bound states in the evolution of $X$, which allows to reduce the complexity of the chemical network into a single Boltzmann equation with an effective cross section. This effective cross section is the output of \codename, and takes into account all relevant processes for the considered gauge theory scenarios: bound-state formation, dissociation, bound state decay and transitions among bound states. In Section~\ref{sec:closed}, we list the closed form expressions of these quantities for in principle arbitrarily highly excited bound states, which enter into the computation of the \codename output. 

\subsection{Effective cross section}
\label{sec:effective}

We consider symmetric scenarios, in which the particle number density, $n_X$, equals the anti-particle ($\bar{X}$) number density. The abundance $n_X$ can change with time due to pair-annihilation of unbound $X\bar X$ pairs (with annihilation cross section times relative velocity $(\sigma v)_\text{ann}$), as well as due to decay of $X\bar X$ bound states. We denote $X\bar X$ bound states by ${\cal B}_i$, labeled by some generic index $i$, encompassing the principal quantum number $n=1,2,3,\dots$ and angular momentum number $0\leq\ell\leq n-1$. Their impact on the evolution of $n_X$ depends on:
 \begin{itemize}
 \item the  cross sections $(\sigma v)_\text{BSF}^i$ of the radiative bound state formation (BSF) process $X+\bar X\to{\cal B}_i+\gamma/g$, where $\gamma$ $(g)$ denotes the $\text{U}(1)$ ($\text{SU}(N)$) gauge boson(s), 
 \item the rates $\Gamma_\text{dec}^i$ of bound state decays ${\cal B}_i\to \gamma\gamma / gg$,
 \item bound-to-bound transition rates $\Gamma_\text{trans}^{i\to j}$. 
 \end{itemize}
 Assuming that the rates of these processes exceed the Hubble rate $H$ during the cosmological epoch relevant for the non-trivial time-evolution of the density $n_X$, the impact of bound states can be captured by a standard Boltzmann equation with an effective, thermally averaged cross section 
 \begin{align}
\dot{n}_{X} + 3 H n_{X} = - \langle \sigma v \rangle_\text{eff} (n^2_{X}-n^2_{X, \text{eq}}) + ...\,.
\end{align}
Here the ellipsis stand for further possible processes that are not related to $X\bar X$ annihilation or bound states, such as e.g.~conversions into different dark sector particle species. The non-relativistic equilibrium number density is $n_{X,\text{eq}}(x)\equiv g_X [mT/(2\pi)]^{3/2}e^{-m/T}$ with $x\equiv m /T$, and $g_X$ are the internal degrees of freedom. 
For example, $g_X=(2s+1)\times N$ if $X$ has spin $s$ and transforms under the fundamental representation of $\text{SU}(N)$. The effective cross section can be split into a part related to annihilation, and one to bound states,
\begin{equation}
  \langle \sigma v \rangle_\text{eff} = \langle(\sigma v)_\text{ann}\rangle + \langle \sigma v \rangle_\text{eff,BSF}\,.
\end{equation}
\codename provides $\svta{}_\text{eff,BSF}$ for classes of models as detailed below. In general, it is given by~\cite{Mitridate:2017izz,Garny:2021qsr,Binder:2021vfo}
\begin{equation}
\label{eq:sigvEff}
\left\langle \sigma \vel \right\rangle_\text{eff, BSF} ~=~ \sum_{i} R_i  \left\langle (\sigma \vel)_\text{BSF}^i \right\rangle,
\end{equation}
where $0\leq R_i \leq 1$ are depletion efficiency factors,
\be
  R_i \equiv  1 - \sum_k (M^{-1})_{ik} \frac{ \Gamma_\text{ion}^k }{ \Gamma_\text{tot}^k }\label{eq:Ri}\,,
\ee
involving the inverse of the transition network matrix
\be
  M_{ik} \equiv \delta_{ik} - \frac{ \Gamma_\text{trans}^{i\to k} }{ \Gamma_\text{tot}^i }\label{eq:Mij}\,,
\ee
where
\be
  \Gamma_\text{tot}^i \equiv \Gamma_\text{ion}^i + \Gamma_\text{dec}^i + \sum_{k\neq i}\Gamma_\text{trans}^{i\to k}\,.\label{eq:Gammai}
\ee
$\Gamma_\text{ion}^i$ is the ionization rate, which relates to the cross section of its inverse process, $(\sigma \vel)_\text{BSF}^i$, by the requirement of detailed balance, known as Milne relation, 
\begin{equation}
  \label{eq:Milne}
  \Gamma_\text{ion}^i = 
  \frac{g_{X}^2}{g_{\mathcal{B}_i}}\left(\frac{m  T}{4\pi}\right)^{3/2}\mathrm{e}^{-|E_i|/T}
   \left\langle (\sigma \vel)_\text{BSF}^i \right\rangle\,,
\end{equation}
where $g_{{\cal B}_i}$ is the number of internal degrees of freedom of the bound state ${\cal B}_i$ with binding energy $E_i$. Finally, the thermal average of the BSF cross section, taking the Bose enhancement factor of the emitted gauge boson into account, reads 
\begin{equation}
\left\langle (\sigma \vel)_\text{BSF}^i \right\rangle \equiv \frac{x^{3/2}}{2\pi^2}\int_0^\infty \mathrm{d}\vel ~\vel^2 e^{-m  v^2/(4T)}(\sigma \vel)_\text{BSF}^i [1+f_B(\omega)]\;, \label{eq:defthermalaverage}
\end{equation}
where $\omega=m v^2/4 +|E_i|$ is the energy of the emitted gauge boson, and $f_B(\omega)=(e^{\omega/T}-1)^{-1}$.

In the following we provide closed-form expressions for the BSF, transition and decay rates in our considered gauge theory scenarios entering in the first release of \codename. 

\subsection{Closed form expressions}
\label{sec:closed}

We collect closed form expressions for all rates entering the effective BSF cross section, including the bound state formation cross sections for arbitrary excited states, bound state decay and transition rates (see~\cite{Garny:2021qsr,Binder:2023ckj,Beneke:2024nxh} and references therein). The models covered by \codename feature bound state formation processes described by dipole transitions, mediated by emission of either a $\text{U}(1)$ or an $\text{SU}(N)$ gauge boson. The expressions collected below apply to both cases. Importantly, for the latter, the colour state in the initial and final state differ from each other. This means in practice that the wavefunctions of scattering and bound states, which enter the BSF cross section, correspond to Coulomb potentials with a different strength for the initial and final state, leading to qualitative differences as compared to the renowned Abelian case.

We consider two cases for the spin of particle $X$, $s=0$ and $s=1/2$. In the latter case, the $X\bar X$ bound states are described by the total spin $S=1,0$ and $m_S=-S,\dots,S$ in addition to $n,\ell$ and $m=-\ell,\dots,\ell$. The corresponding multiplicity factors for bound states ${\cal B}_{n\ell S}$ are $g_{{\cal B}_{n\ell S}}=(2\ell+1)(2S+1)$. For the leading bound state formation and transition rates considered here, the dipole interaction is spin-independent, and the only spin-dependence enters via the bound state decay rate, apart from multiplicity factors due to the number of states. Consequently, for $s=1/2$, the effective BSF cross section can be split into a sum of a contribution involving only the set of ${\cal B}_{n\ell, S=1}$ states and one from ${\cal B}_{n\ell, S=0}$ states, respectively. Since the decay rate of the $S=1$ states is strongly suppressed compared to those for $S=0$~\cite{Binder:2020efn,Binder:2023ckj}, also the contribution to the effective BSF cross section from $S=1$ states is suppressed, and we focus on $S=0$ states below.\footnote{The only exception are models for which $X$ is a colour neutral but electrically charged fermionic mediator. In that case, the $S=1$ bound states can decay into pairs of light SM fermions via a photon in the $s$-channel, with decay rate related to that of the $S=0$ bound states as given in Eq.~(5.4) in~\cite{Binder:2020efn}. The large number of light, electrically charged fermions in the SM makes the $S=1$ decay rate comparable to that of the $S=0$ state in that particular case. We therefore include the contribution from both $S=1$ and $S=0$ states to the effective BSF cross section for the class of models referred to as \texttt{QED} in \codename, while for all other cases only the $S=0$ contribution is included.} 
Therefore, we hide the spin quantum number $S$ in the following and adopt the convention that all quantities apply to the bound states ${\cal B}_{n\ell}\equiv {\cal B}_{n\ell,S=0}$ with arbitrary $n,\ell$ and $S=0$, with multiplicity factor $g_{\mathcal{B}_{n\ell}}=2\ell+1$.

The radiative BSF process via gauge boson emission is described as a dipole transition between unbound scattering states with relative momentum $p=m v/2$ and angular momenta $\ell'=|\ell\pm 1|$, and summing over all $g_{\mathcal{B}_{n\ell}}=2\ell+1$ bound states with principal quantum number $n$ and angular momentum $\ell$,
\begin{equation}
  (\sigma \vel)_\text{BSF}^{n\ell} =
  (\sigma v)_{p \lp \to n \el}^\text{dipole}\Big|_{\el'=\el-1} + (\sigma v)_{p \lp \to n \el}^\text{dipole}\Big|_{\el'=\el+1} \,,
\end{equation}
where
\begin{equation}\label{eq:svBSFdipole}
(\sigma v)_{p \lp \to n \el}^\text{dipole} = 4\pi \alpha_{\BSF} \,\frac{4\omega^{3}}{9} \times I_{A,\text{dipole}}^\BSF \times I_{R,\text{dipole}}^\BSF \times\frac{1}{(2s+1)^2}
\,,
\end{equation}
with squared angular matrix element
\be
  I_{A,\text{dipole}}^\BSF = 
3(2\el+1)(2\lp+1) 
  \begin{pmatrix}
    1 & \el & \lp\\
    0 & 0 & 0
  \end{pmatrix}^2
\,,
\ee
and radial part $I_{R,\text{dipole}}^\BSF$. The BSF process is in general dependent on three distinct effective coupling parameters, that are all related to the underlying gauge coupling:
\begin{itemize}
\item $\alpha_\text{BSF}$ related to the emission of the gauge boson,
\item $\alpha_s$ related to the effective strength of the potential $V_s(r)=-\alpha_s/r$ of the initial scattering state,
\item $\alpha_b$ related to the effective strength of the potential $V_b(r)=-\alpha_b/r$ of the final bound state\,.
\end{itemize}
For a $\text{U}(1)$ gauge interaction responsible for the potentials as well as the emitted gauge boson, all of these are equal to the gauge coupling strength multiplied with the squared charge $Q_X$ of $X$, 
\begin{equation}
\alpha_\text{BSF}\big|_{\text{U}(1)} = \alpha_s\big|_{\text{U}(1)} = \alpha_b\big|_{\text{U}(1)} = Q_X^2\times \alpha\big|_{\text{U}(1)}\,.
\end{equation}
For $\text{SU}(N)$, 
\begin{eqnarray}
  \alpha_\text{BSF}\big|_{\text{SU}(N)} &=& \frac{C_F}{N^2}\times \alpha(\mu_\text{BSF})\big|_{\text{SU}(N)} \to \frac4{27}\times \alpha(\mu_\text{BSF})\big|_{\text{SU}(3)}\,,\nonumber\\
  \alpha_s\big|_{\text{SU}(N)} &=& (C_F-C_A/2) \alpha(\mu_s)\big|_{\text{SU}(N)} \to -\frac16\times \alpha(\mu_s)\big|_{\text{SU}(3)}\,,\nonumber\\
  \alpha_b\big|_{\text{SU}(N)} &=& C_F\alpha(\mu_b)\big|_{\text{SU}(N)} \to \frac43\times \alpha(\mu_b)\big|_{\text{SU}(3)}\,,
\label{eq:alphab}
\end{eqnarray}
where $C_F=(N^2-1)/(2N), C_A=N$, and the last expressions correspond to $N=3$. For the case of QCD, we use the known scale-dependence $\alpha(\mu)\big|_\text{QCD}$ (see Sec.~\ref{sec:impl} for details) and evaluate the running coupling at the respective typical momentum scales $\mu_\text{BSF}=\omega$, $\mu_s=p=m v/2$ and $\mu_b=p_n\equiv m \alpha_b/(2n)$. 
For a dark sector $\text{SU}(3)$, the running depends on the unknown dark sector particle content, and we use a constant value $\alpha\big|_\texttt{dQCD}$ entering all three effective coupling strengths for definiteness.
We use the notations
\begin{equation}
  \zeta_s\equiv \alpha_s/v,\quad
  \zeta_n\equiv \alpha_b/(nv),\quad
  \kappa\equiv \alpha_s/\alpha_b\,.
\end{equation}
The binding energy reads $|E_n|=m \alpha_b^2/(4n^2)$. The radial part of the squared dipole matrix element entering the BSF cross section~\eqref{eq:svBSFdipole} is given by~\cite{Beneke:2024nxh}
\begin{align}
\label{eq:IRBSF}
I_{R,\text{dipole}}^\BSF &~=~ 
\frac{2^{4 \el+2} \zbb^{2 \el+3} }{ (\mu v)^{5}\lr 1+\zbb^2\rr^{2\el+4}}
\frac{\Gamma(\lp+1)^2  \Gamma (n+\el+1) }{n \Gamma (2 \el+2)^2 \Gamma (n-\el)}
S_{\lp}(\zs)\,\E^{-4\zs \gbb }\nonumber
\\&\qquad\nonumber\times
 \left|\frac{ 1-\E^{2\I\lr2(n-\el)\gbb-\gamma_F-\gamma_R\rr}}
{ n\kappa\zbb\lr\zbb^2- 1+\frac{2}{\kappa} \rr} \right|^2
 \left|F_+(0)\right| ^2\left|R^\text{dipole}_{\lp-\el}\right|^2
\,,\nonumber\\
F_+(0) &~\equiv~ {_2F_1}\lr -n+\el+1 ,\, \I\zs + \el ,\, 2\el+2 ,\, \frac{-4\I\zbb}{(\zbb-\I)^2} \rr
\,,\\
\label{eq:SE}
 S_{\lp}(\zs)  &~\equiv~ 
\E^{\pi\zs}\frac{\left| \Gamma\lr 1+\lp-\I \zs \rr \right|^2}{\Gamma(1+\lp)^2}
\,,\nonumber\\
R^\text{dipole}_{1} &~\equiv~ 
\zs(1+\zbb^2) +n\zbb(1-\kappa)\lr 2+2\I n\zbb(1-\kappa)+(\el+1)(1+\zbb^2)  \rr 
\,,\nonumber\\
R^\text{dipole}_{-1} &~\equiv~ 
\left[    \zs(1+\zbb^2) +n\zbb(1-\kappa)\lr 2 +2\I n\zbb(1-\kappa)-\el(1+\zbb^2) \rr    \right]\nonumber\\
    & \qquad \times \lr \el-\I\zs\rr \lr1+\el-\I\zs \rr
\,.\nonumber
\end{align}
Here $\gbb$, $\gamma_F$ and $\gamma_R$ denote the complex phases of $(\I+\zbb)$, $F_+(0)$ and $R^\text{dipole}_{\lp-\el}$, respectively. 

For bound-state decays, we consider the leading contributions from $s$-states (see, e.g.~\cite{Binder:2023ckj})
\begin{align}
\Gamma_\text{dec,\,\text{SU}(3)}^{n\ell\to gg} &= \delta_{\ell,0}\,\frac{m\,C_F}{4n^3} \lr\alpha_\text{SU(3)}(\mu_h)\rr^2  
 \alpha_b^3 \times\frac{2s+1}{2}\,,
\\
\Gamma_\text{dec,\,\text{U}(1)}^{n\ell\to \gamma \gamma} &= \delta_{\ell,0}\,\frac{m\,\alpha^5}{2n^3} \times\frac{2s+1}{2}\,.
\end{align}
Where running couplings are included, the hard scale is evaluated at $\mu_h=m$ for the short distance process of the decay, while $\alpha_b$, cf.~\eq~(\ref{eq:alphab}), arises from the bound-state wavefunction.

Lastly, we consider bound-to-bound transitions by absorption or emission of a $\text{U}(1)$ gauge boson, corresponding to excitations $n'\ell'+\gamma\to n\ell$ or de-excitations $n'\ell'\to n\ell+\gamma$. We note that even when $X$ is charged under $\text{SU}(N)$, transitions cannot occur via absorption or emission of a non-Abelian gauge boson due to colour conservation. Thus, transitions are relevant if
\begin{itemize}
\item $X$ is charged under $\text{U}(1)$ but neutral under $\text{SU}(3)$. In this case the $\text{U}(1)$ coupling determines both the bound state wavefunctions as well as the transition strength.
\item $X$ is charged under $\text{U}(1)$ and $\text{SU}(3)$. In this case we assume that bound state wavefunctions are dominated by the $\text{SU}(3)$ interaction, while the $\text{U}(1)$ is responsible for the transition process. This applies in particular to coloured $t$-channel mediator particles $X$.
\end{itemize}
In both cases, the de-excitation rates ($n'>n$) can in general be written as (see, e.g.~\cite{Binder:2023ckj})
\begin{align}
    \Gamma_\text{trans}^{n'\lp\to n\el} = \frac{4 Q^2_{X}\,\alpha_{\text{U}(1)}}{3} (2\ell+1)(\omega_{n'n})^3 \,\left|\bra{\psi_{n^\prime\ell^\prime}} \mathbf{r} \ket{\psi_{n \ell}}\right|^2\,[1+f_B(\omega_{n'n})], \label{eq:dex}
\end{align}
where $\omega_{n'n}=|E_n|-|E_{n'}|>0$ is the energy difference, $\alpha_{\text{U}(1)}$ the coupling strength parameter of the $\text{U}(1)$ interaction responsible for the transition, and $Q_X$ the $\text{U}(1)$ charge of $X$. Note that we do not include running of the $\text{U}(1)$ coupling strength. The excitation rates are computed in terms of those for de-excitation using the detailed balance relation 
\begin{equation}
\Gamma_\text{trans}^{n\ell\to n'\ell'}=\Gamma_\text{trans}^{n'\ell'\to n\ell}\,\frac{g_{\mathcal{B}_{n'\lp}}}{g_{\mathcal{B}_{n\ell}}}\,e^{- \omega_{n'n}/T}.
\end{equation} 
The matrix elements for the contributions with angular momentum increase ($\lp=\el-1$) and decrease ($\lp=\el+1$) are given by 
\begin{align}
    \left|\bra{\psi_{n^\prime\ell^\prime}} \mathbf{r} \ket{\psi_{n \ell}}\right|^2 = \frac{\ell'\delta_{\ell',\ell+1}+\ell\delta_{\ell,\ell'+1}}{(2\ell+1)(2\ell'+1)}\,|I_{R,\text{trans}}|^2\,,
\end{align}
where
\begin{align}
\left|I_{R,\text{trans}}^{\lp = \el +1}\right|& = N_{n,\el}(p_{n})N_{n^\prime,\el+1}(p_{n'}) \times \label{eq:trlp1}
    \\\nonumber  & 2^{-1} (\el+1)(2\el+3)\Gamma(2\el+2) (1-z)^{\frac{n^\prime -n}{2}}\frac{z^{\el+1}}{p_{n'}^2p_{n}}\times
    \\\nonumber  &\left|
    F_{+,t}(0) \left(\frac{(l-n'+1) (n+n') (n' p_{n'}-n p_n)}{n' (n'-n) \left(p_n^2-p_{n'}^2\right)}-\frac{n p_n-n' p_{n'}+p_{n'}}{(p_n-p_{n'})^2}\right)\right.
    \\\nonumber &\left.+F_{+,t}(2) \left(\frac{-n p_n+n' p_{n'}+p_{n'}}{(p_n+p_{n'})^2}-\frac{(l+n'+1) (n'-n) (n' p_{n'}-n p_n)}{n' (n+n') \left(p_n^2-p_{n'}^2\right)}\right)
    \right|
\\
\left|I_{R,\text{trans}}^{\lp = \el - 1}\right| & = N_{n,\ell}(p_{n})N_{n^\prime,\el-1}(p_{n'}) \times \label{eq:trlm1}
    \\\nonumber  & 2^{-1} \el(2\el+1)\Gamma(2\el) (1-z)^{\frac{n^\prime -n}{2}} \frac{z^\el}{p_{n}^2 p_{n'}}\times 
    \\\nonumber &\left|
    F_{-,t}(-2) \left(\frac{n' p_{n'}-(n+1) p_n}{(p_n-p_{n'})^2}-\frac{(l+n) (n+n') (n p_n-n'   p_{n'})}{n (n-n') \left(p_n^2-p_{n'}^2\right)}\right)\right.
    \\\nonumber&\left.+F_{-,t}(0) \left(\frac{(l-n) (n-n')   (n p_n-n' p_{n'})}{n (n+n') \left(p_n^2-p_{n'}^2\right)}+\frac{-n p_n+n'   p_{n'}+p_n}{(p_n+p_{n'})^2}\right)
\right|
\end{align}
Here we use the abbreviations\footnote{Our treatment of the numerical evaluation of hypergeometric ${}_2F_1$ functions occurring in transition rates for large $n\gtrsim 10$ and $z>0.7$ is explained in Appendix A.2 of \cite{Binder:2023ckj}.} 
\begin{align}
z&\equiv \frac{4  p_{n} p_{n'} }{(p_{n}+p_{n'})^2}<1\,,\nonumber\\
N_{n,\el}(p) &\equiv \frac{p^{3/2}}{\sqrt{n}} \frac{2}{(2\el+1)!} \left( \frac{(n+\el)!}{(n-\el-1)!}\right)^{1/2}\,,\nonumber\\
F_{+,t}(k) &\equiv {}_2 F_1\left(\ell-n+1,\ell+n'+k,2 \ell+2,z\right)\,,\nonumber
\\
F_{-,t}(k) &\equiv {}_2F_1\left(\ell-n+1+k,\ell+n',2 \ell,z\right)\,.
\end{align}
We note that the bound state coupling strength enters via $p_n=m \alpha_b/(2n)$ and $p_{n'}=m \alpha_b'/(2n')$, respectively. When considering a running coupling as described above, we stress that $\alpha_b$ depends on $n$, and denote the coupling evaluated for level $n'$ by $\alpha_b'$.

When neglecting the running of the gauge coupling, the effective coupling strength parameters for the initial and final bound state coincide (i.e. $\alpha_b=\alpha_b'$). In this case, the rates can be simplified considerably and read 
\begin{align}
\notag
\left|I_{R,\text{trans}}^{\lp = \el +1}\right| =& \,
4(\ab)^2\frac{z^{2\el} |1-z|^{1+n-n'}}{ \Gamma(2\el+2)^2} \,\frac{\Gamma(n+\el+1)}{\Gamma(n-\el)}\frac{\Gamma(n'+\el+2)}{\Gamma(n'-\el-1)}
\\&~\times 
\lr \frac{F_{+,\text{t}}(0)}{(n'-n)^2}-\frac{F_{+,\text{t}}(2)}{(n'+n)^2}\rr^2
,\\
\notag
\left|I_{R,\text{trans}}^{\lp = \el -1}\right| =& \,
4(\ab)^2\frac{z^{2\lp} |1-z|^{1+n'-n}}{ \Gamma(2\lp+2)^2} \,\frac{\Gamma(n+\el+1)}{\Gamma(n-\el)}\frac{\Gamma(n'+\el)}{\Gamma(n'-\el+1)}
\\&~\times 
\lr \frac{F_{-,\text{t}}(0)}{(n'+n)^2}-\frac{F_{-,\text{t}}(2)}{(n'-n)^2}\rr^2
,
\end{align}
with $z\to 4 n' n/(n'+n)^2$ 
when neglecting running.

\begin{figure}[t]
\begin{center}
\includegraphics[width=0.8\textwidth]{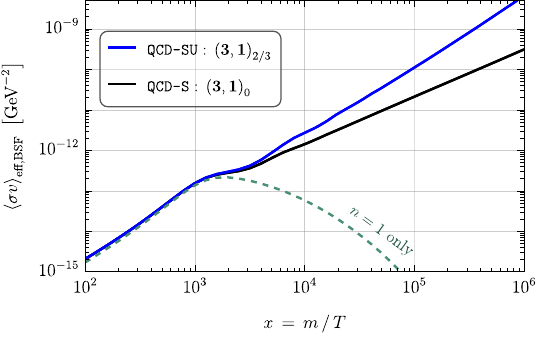}
\caption{\label{fig:ad} Bound state contribution $\langle\sigma v\rangle_\text{eff,BSF}$ to the thermally averaged effective cross section for a scalar $X$ with quantum numbers identical to those of the right-handed up-type quarks (`stop'-like, see \tab~\ref{tab:model-coverage}) provided by \codename.  The blue line shows the full result when taking bound state formation/ionization, bound state decays and transitions among bound states into account. The black line corresponds to the case when neglecting transitions (technically applicable to a colour-charged but electrically neutral scalar particle). The dashed line shows the contribution from the ground state only ($n=1$) for illustration. All lines correspond to $m =10^6$\,GeV and use the running QCD as well as QED couplings.
}
\end{center}
\end{figure}

\medskip

In Fig.~\ref{fig:ad}, we show an example for the thermally averaged effective cross section $\langle \sigma v \rangle_\text{eff,BSF}$ computed by \codename for a coloured scalar mediator $X$ with SM gauge quantum numbers identical to those of the top quark (``stop-like''). While BSF and decay processes are dominated by QCD interactions, the electromagnetic interaction gives rise to transitions among bound states. We also show, for comparison, the case when computing the effective cross section while neglecting transitions, as well as the contribution from the ground state only. Excited states dominate the effective cross section for $x=m /T\gtrsim m /|E_1|$. At larger $x$, those excited bound states with binding energies $|E_n|$ of comparable magnitude than $T$ provide the main contributions to $\langle \sigma v \rangle_\text{eff,BSF}$, being a characteristic feature of BSF mediated by non-Abelian interactions. As discussed further below, we note that within the range of $x$ provided by \codename, the running gauge coupling remains within the perturbative regime and the BSF cross sections satisfy perturbative partial wave unitarity limits. Furthermore, the amount of excited states ($n\leq 100$) that is explicitly included in \codename is sufficient to obtain converged results within the shown $x$ range (see Sec.~\ref{sec:impl} for details).

\section{Rescaling of the effective cross section}
\label{sec:rescale}
The effective BSF cross section in Eq.~\eqref{eq:sigvEff} is a function of the temperature parameter $x$ for a set of model parameters, such as the mass $m$ and coupling parameters of the species:
\begin{align}
\langle \sigma v \rangle_{\mathrm{eff,BSF}}
= \langle \sigma v \rangle_{\mathrm{eff,BSF}}\left(x ; \{\text{model parameters}\}\right) .
\label{eq:EFFBSFDEP}
\end{align}
A straightforward tabulation of  $\langle \sigma v \rangle_{\mathrm{eff,BSF}}$ would therefore require multidimensional grids in all relevant model parameters in addition to the temperature parameter $x$. This naturally raises the question under which conditions effective BSF cross sections corresponding to different choices of model parameters are related.

We demonstrate that such relations indeed exist, so that the information contained in a single tabulation can be reused to cover a much wider range of parameter values. Specifically, we show that under the most favorable conditions, Eq.~\eqref{eq:EFFBSFDEP} needs to be evaluated only once for a single reference set of parameters over a sufficiently broad range in $x$. 
In such optimal cases, the dimension of the effective BSF cross section is minimal and reduces to a one-dimensional table in $x$. Other model parameter values can then be obtained through simple rescalings of $x$ and corresponding multiplicative prefactors. \codename provides the minimal dimension output, thus, these reductions are essential.

\subsection{Frozen coupling}
\label{sec:frozen}

We start the discussion on such extreme dimensional reductions by considering the condition of a \emph{frozen} coupling at the relevant scales introduced in Section~\ref{sec:closed}:
\begin{align}
\alpha = \alpha (\mu_\text{h})=\alpha (\mu_s)=\alpha (\mu_b)= \alpha (\mu_{\text{BSF}})\;.
\end{align}
This is indeed the case for dark QED, where the dark photon only couples to the species $X$. Hence the coupling parameter $\alpha$ runs only for scales larger than the hard scale $\mu_h=m$. In such a case, it turns out that the effective BSF cross section for gauge theories as generically given in Section \ref{sec:closed} follows a simple rescaling law in the mass $m$:
\begin{align}
\langle \sigma v \rangle_\text{eff,BSF}(x;m^\prime,\alpha)= \left(\frac{m}{m^\prime}\right)^2  \times \langle \sigma v \rangle_\text{eff,BSF}(x;m,\alpha)\;.\label{eq:mscaling}
\end{align}
This is due to the fact that the rates, $\Gamma_\text{dec}^i$, $\Gamma_\text{trans}^{i\rightarrow k}$, and $\Gamma_\text{ion}^{i}$, are all proportional to $m$, such that the efficiency factors $R_i$, which include only ratios of these rates, are independent of $m$. Consequently, the mass dependence of $\langle \sigma v \rangle_\text{eff,BSF}$ is entirely set by $\langle (\sigma v)^i_\BSF\rangle$, which is proportional to $m^{-2}$. This explains Eq.~\eqref{eq:mscaling}.

Having established a scaling law in $m$ for fixed $\alpha$, we turn to a scaling law in $\alpha$ for fixed $m$. We shall drop the parametric dependence on $m$ in the following for notational simplicity. We find that for frozen couplings, the parametric dependence of the BSF cross section on $\alpha$ and $v$ can be written as:
\begin{align}
(\sigma v)^i_\BSF = \alpha^{2L}f^i(\alpha/\vel) \;,   
\end{align}
with $L=1$ for dipole interactions on which we focus here. The thermal average of $(\sigma v)^i_\BSF$ in Eq.~\eqref{eq:defthermalaverage} involves the product of the Bose-enhancement factor for the emitted gauge field and the thermal distribution in relative velocity space. One can recognize that this product only depends on $xv^2$ and the ratio $\alpha/v$. Consequently, the parametric dependence of the thermally averaged BSF cross section can be written as
\begin{align}
\langle (\sigma v)^i_\BSF\rangle (x;\alpha) = x^{3/2}\int_0^\infty \mathrm{d}v ~ v^2 [ {\alpha}^2 g^i(x\vel^2,\alpha/\vel) ] \,,
\end{align}
where the function $g^i$ depends on the product $x v^2$, and not separately on $x$. Due to this and the integral limits, the thermally averaged BSF cross sections for different $\alpha$ values are all related by a rescaling of the temperature parameter $x$ and a multiplicative factor. Concretely, introducing the rescaling factor
\begin{align}
r=\frac{\alpha^\prime}{\alpha}
\end{align}
and
transforming the integration variable as $\vel \to \vel' = \vel / r$ and the temperature parameter as $x^\prime = x r^2$, one finds:
\begin{align}
\langle (\sigma v)^i_\BSF\rangle (x;\alpha^\prime) = r^2 \times \langle (\sigma v)^i_\BSF\rangle \left(x^\prime ; \alpha \right)\;.
\end{align}
Inserting this into the Milne relation in Eq.~\eqref{eq:Milne} yields the rescaling law of the ionization rates:
\begin{align}
\Gamma_\text{ion}^i (x;\alpha^\prime) = r^5 \times \Gamma_\text{ion}^i (x^\prime;\alpha)\;.
\end{align}
Furthermore, the parametric dependence of the transition rates can be written as:
\begin{align}
\Gamma_\text{trans}^{i\rightarrow k}(x;\alpha) = \alpha^5 h^{i\rightarrow k}(x \alpha^2) \;.
\end{align}
This can be verified from the general dipole expressions for de-excitation in Eq.~\eqref{eq:dex}, by counting each bound-state momentum as $p_{n^{(\prime)}}\propto \alpha$ and $\omega \propto \alpha^2$. Consequently, the rescaling law of bound-to-bound transition rates is given by 
\begin{align}
 \Gamma_\text{trans}^{i\rightarrow k}(x;\alpha^\prime) = r^5 \times \Gamma_\text{trans}^{i\rightarrow k}(x^\prime;\alpha)\;.
\end{align}
Last, we consider the bound state decay rates, which are temperature independent. For the leading $s$-wave (and spin-singlet) decay rates considered in this work, we have the rescaling law \begin{align}
\Gamma_\text{dec}(\alpha^\prime) = r^5 \times \Gamma_\text{dec}(\alpha).
\end{align} As the depletion efficiency factor $R_i$  involves only certain ratios of ionization, bound-to-bound and decay rates, the common $r^5$ factors cancel, leading to the rescaling law:
\begin{align}
R_i(x;\alpha^\prime)= R_i(x^\prime; \alpha)\,.
\end{align}
Thus, ultimately, the effective BSF cross section satisfies the  rescaling law in $\alpha$:
\begin{align}
\langle \sigma v \rangle_\text{eff,BSF}(x;\alpha^\prime)= r^2  \times \langle \sigma v \rangle_\text{eff,BSF}(x^\prime;\alpha) \;.\label{eq:alphascaling}
\end{align}
Notice that this is an exact analytic expression for dipole transitions, when neglecting the running of the coupling. Combining mass and coupling rescaling laws in Eq.~\eqref{eq:mscaling} and Eq.~\eqref{eq:alphascaling}, respectively, we finally obtain under the condition of \emph{frozen} couplings:
\begin{align}
\langle \sigma v \rangle_\text{eff,BSF}^\text{rescaled}(x;m,\alpha)= \left(\frac{m_0 \alpha}{m \alpha_0}\right)^2  \times \langle \sigma v \rangle_\text{eff,BSF}(x \times (\alpha/\alpha_0)^2; m_0,\alpha_0)\;. \label{eq:scalinglaw}
\end{align}
Once the effective BSF cross section on the right hand side is computed for a single reference mass $m_0$ and coupling value $\alpha_0$ in a sufficiently large temperature range, effective BSF cross sections for all other parameter values $m$ and $\alpha$ follow from this law. This allows us to describe several of the cases covered in \codename with one-dimensional tables in $x$ (see Table~\ref{tab:model-coverage} in Sec.~\ref{sec:modelcov}). 

\subsection{Running coupling}

Suppose that the effective BSF cross section involves multiple but known couplings with given runnings. In such a case we can reduce the dimensionality to always two-dimensional tables, $\svta{}_\text{eff,\BSF}=\svta{}_\text{eff,\BSF}(x;m)$, without explicit dependence on couplings. This is because the open parameter $m$ is the only unknown in all relevant scales $\mu \in \{ \mu_h, \mu_s, \mu_b, \mu_{\text{BSF}} \}$ entering the evaluation of the couplings. 
Prime examples are those where the species is charged under Standard Model gauge groups, see  Table~\ref{tab:model-coverage} in Sec.~\ref{sec:modelcov}.

Finding rescaling laws for running couplings is more involved. The exact relations discussed in Section~\ref{sec:frozen} rely on scale independence and therefore break down once running is included. Nevertheless, it is worth asking whether effective BSF cross sections computed with frozen couplings can still provide useful \emph{approximate} descriptions when running effects are present.

To explore this possibility, we consider the limit in which bound-to-bound transitions are absent, such that the depletion efficiency factor reduces to:
\begin{align}
R_i =  \frac{1}{1+(\Gamma^i_\text{ion}/\Gamma^i_\text{dec})} \;.
\end{align}
This is indeed the case in a pure SU($N$) theory (such as dark QCD) in the approximation we are working in. Furthermore, let us assume that the running of the coupling is given, denoted by $\alpha(\mu)$. As before, the parametric dependence of the full effective cross section, including the running of the coupling, can be written as $\svta{}_\text{eff,\BSF}=\svta{}_\text{eff,\BSF}(x;m)$ without explicit dependence on $\alpha$. 

Let us consider the size of a typical relative momentum $p\sim \sqrt{mT}$ of a scattering state, which defines the potential scale:
\begin{align}
\mu_p \equiv \sqrt{mT} \;.
\end{align}
Notice that $\mu_p \sim \mu_s \sim \mu_b$ at temperatures around the binding energy of each bound state. We empirically find that the effects of a running coupling in full $\svta{}_\text{eff,\BSF}$ can be captured to good approximation by evaluating the frozen-coupling result, $\svta{}_\text{eff,BSF}^\text{rescaled}$ in Eq.~\eqref{eq:scalinglaw}, close to this potential scale:
\begin{align} 
\svta{}_\text{eff,\BSF}(x;m)\approx \svta{}_\text{eff,BSF}^\text{rescaled}\lr x;m,\alpha(C \mu_p)\rr \;. \end{align}
where $C$ is an $\mathcal{O}(1)$ number, which can be optimised to the problem at hand.

\begin{figure}[t]
\begin{center}
\includegraphics[width=0.66\textwidth]{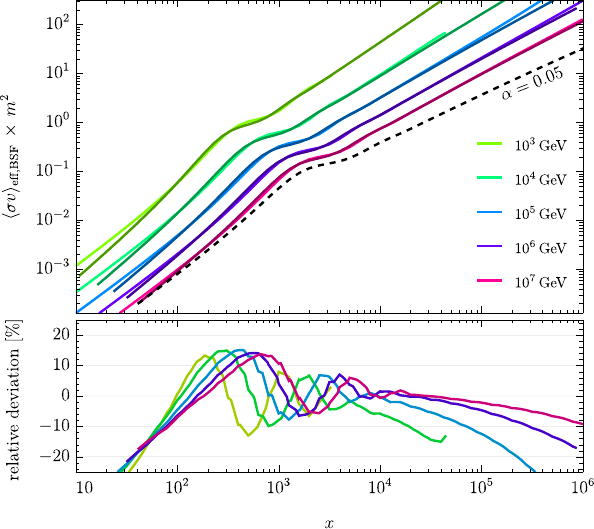}
\caption{\label{fig:Rescaling}
Comparison of \texttt{QCD-S} models and approximately rescaled \texttt{dQCD-S} models (darker lines) where the SU(3) gauge coupling at every $x$ is chosen at the potential scale $\alpha_\text{s}=\alpha_\text{s}(\sqrt{2m T})$. \emph{Upper panel}: $m^2 \svta{}_\text{eff,\BSF}$ plotted over temperature for different masses (different colours). The black dashed line shows the result for setting $\alpha=0.05$ at all $x$ which approximates $\alpha_\text{QCD}(10^7\,\text{GeV})=0.0460$. Note that choosing a different $\alpha$ here merely shifts the drawn curve around but affects neither its shape nor slope. \emph{Lower panel:} Relative deviation of the approximately rescaled \texttt{dQCD-S} to \texttt{QCD-S} in percent. 
}
\end{center}
\end{figure}

Figure~\ref{fig:Rescaling} illustrates this approximation for a representative example with sizable running effects. The figure compares the effective BSF cross section obtained from a full computation including running couplings with the corresponding frozen-coupling result evaluated at the potential scale. Across almost the entire perturbative temperature range, the agreement is at the level of $\lesssim 15\%$, demonstrating that this simple prescription captures the dominant impact of running couplings on the effective BSF cross section.

We leave a more systematic study of improved approximate rescaling schemes for future work. For the purposes of this paper, the example shown here serves to illustrate that the rescaling framework developed above can remain useful beyond the strictly non-running case, even though the relations cease to be exact.

\section{Coverage, implementation,  and limitations}
\label{sec:impl}

\subsection{Model coverage}
\label{sec:modelcov}

\codename includes a set of models that covers a broad range of phenomenologically relevant BSF scenarios. The models are summarised in Table~\ref{tab:model-coverage} and are organised according to the underlying gauge interactions (SM QCD, dark QCD, SM QED, and dark QED), the spin of the constituents and the presence or absence of radiative transitions.

The guiding principle behind this selection is to balance phenomenological relevance with computational efficiency. For instance, the cases involving bound states of new particles carrying SM gauge charges find a natural application in so-called $t$-channel mediator dark matter models. In these models, the mediator shares the gauge quantum numbers of one of the SM fermions -- a quark or lepton -- allowing for a Yukawa-type interaction between that fermion, the mediator, and the dark matter particle; see Ref.~\cite{Arina:2025zpi} for a comprehensive review. Bound-state effects can strongly enhance mediator-pair annihilation rates, which is highly relevant for a variety of viable dark matter production mechanisms within these models, such as coannihilation, conversion-driven freeze-out, and superWIMP production.

\begin{table}[t]
\centering
\renewcommand{\arraystretch}{1.25}
\footnotesize
\begin{tabular}{c c c c c c c}
\hline\hline
 SU(3) rep.\!\! & $|Q|$ & spin & 
\!our model name\!\! &  maps to models in \cite{Arina:2025zpi} & rescaling & parameters \\
\hline
$\mathbf{3}$ & $2/3$  & $0$   & \texttt{QCD-SU} & \{\texttt{S3M},\texttt{S3D}\}\texttt{\_}\{\texttt{uR},\texttt{cR},\texttt{tR}\} & -- & $x,m$ \\
$\mathbf{3}$ & $1/3$ & $0$   & \texttt{QCD-SD} & \{\texttt{S3M},\texttt{S3D}\}\texttt{\_}\{\texttt{dR},\texttt{sR},\texttt{bR}\} & -- & $x,m$ \\
$\mathbf{3}$ & 0 & $0$   & \texttt{QCD-S} & -- & -- & $x,m$ \\
$\mathbf{3}$ & $2/3$  & $1/2$ & \texttt{QCD-FU} & \{\texttt{F3S},\texttt{F3C},\texttt{F3V},\texttt{F3W}\}\texttt{\_}\{\texttt{uR},\texttt{cR},\texttt{tR}\} & -- & $x,m$ \\
$\mathbf{3}$ & $1/3$ & $1/2$ & \texttt{QCD-FD} & \{\texttt{F3S},\texttt{F3C},\texttt{F3V},\texttt{F3W}\}\texttt{\_}\{\texttt{dR},\texttt{sR},\texttt{bR}\} & -- & $x,m$ \\
$\mathbf{3}$ & $0$ & $1/2$ & \texttt{QCD-F} & -- & -- & $x,m$ \\
$\mathbf{3}$ & 0 & $0$   & \texttt{dQCD-S} & -- & \checkmark & $x,m,\alpha$ \\
$\mathbf{3}$ & $0$ & $1/2$ & \texttt{dQCD-F} & -- & \checkmark & $x,m,\alpha$ \\
\hline
$\mathbf{1}$ & $1$   & $0$   & \texttt{QED-S}  & \{\texttt{S3M},\texttt{S3D}\}\texttt{\_}\{\texttt{eR},\texttt{muR},\texttt{taR}\} & \checkmark & $x,m$ \\
$\mathbf{1}$ & $1$   & $1/2$ & \texttt{QED-F}  & \{\texttt{F3S},\texttt{F3C},\texttt{F3V},\texttt{F3W}\}\texttt{\_}\{\texttt{eR},\texttt{muR},\texttt{taR}\}
 & \checkmark & $x,m$ \\
$\mathbf{1}$ & $1$   & $0$   & \texttt{dQED-S}  & -- & \checkmark & $x,m,\alpha$ \\
$\mathbf{1}$ & $1$   & $1/2$ & \texttt{dQED-F}  & -- & \checkmark & $x,m,\alpha$ \\
$\mathbf{1}$ & $1$   & $0$   & \texttt{dQED-SnoTr}  & -- & \checkmark & $x,m,\alpha$ \\
$\mathbf{1}$ & $1$   & $1/2$ & \texttt{dQED-FnoTr}  & -- & \checkmark & $x,m,\alpha$ \\
\hline\hline
\end{tabular}
\caption{
Overview of the covered models. The first three columns show the particle quantum numbers (QCD representation, electric charge, and spin), the forth column the \codename model name, and the fifth column the corresponding $t$-channel mediator models it applies to as classified in the review \cite{Arina:2025zpi}. The sixth column indicates whether we employ the rescaling described in Sec.~\ref{sec:rescale} and last one displays the free parameter of the model. The models \texttt{dQED-SnoTr} and \texttt{dQED-FnoTr} are the same as \texttt{dQED-S} and \texttt{dQED-F}, respectively, but neglecting bound-to-bound transitions.  For \texttt{QED-S} and \texttt{QED-F}, the electromagnetic coupling is used by default;
see Appendix~\ref{app:usage} for details.
}
\label{tab:model-coverage}
\end{table}

For convenience, we list the corresponding 
$t$-channel mediator models -- following the nomenclature of Ref.~\cite{Arina:2025zpi} -- to which our results apply in the fifth column of Table~\ref{tab:model-coverage}.  Note that our results are also applicable to a broad class of flavoured dark matter scenarios in which the mediator carries the same SM charges as specified in Table~\ref{tab:model-coverage}; see Ref.~\cite{Belfatto:2025ids} for a recent classification.  They further extend to models with multiple 
$t$-channel mediators of different flavour, provided that the decay of bound states composed of mediators of different flavour is sufficiently suppressed to render their contribution negligible, as is typically implied by flavour constraints. In these cases, the effective cross section depends only on the long-range gauge interactions of the mediator pair, not on the flavour structure of the dark matter–SM coupling.

Beyond dark matter models, our results are relevant for other beyond-SM scenarios in the early Universe, such as the appearance of long-lived charged particles in the context of baryogenesis. An example is provided in Ref.~\cite{Heeck:2023soj} realizing Dirac leptogenesis in a regime that shares similarities with the superWIMP scenario. Both quark-philic and leptophilic models are covered by our tool. 

For the models involving SM gauge interactions, the coupling strengths are known and the tabulated grid involves the two parameters $m$ and $x$. In the case of SM QCD, a running coupling $\alpha_\text{s}$ is used, see Sec.~\ref{sec:numimpl}, while we neglect running for the QED coupling strength, $\alpha_\text{em}$. In the former case, we provide two different prescriptions for $\alpha_\text{s}$ when evaluated below a scale of $1\,\text{GeV}$. The two choices are setting $\alpha_\text{s}(\mu\!<\!1\,\text{GeV})=0$ (cutoff) or $\alpha_\text{s}(\mu\!<\!1\,\text{GeV})=\alpha_\text{s}(\mu\!=\!1\,\text{GeV})\approx0.5$ (plateau). Note that the former is used by default when no extension is provided.

Beyond SM gauge interactions, we also provide results for particles charged under a generic $\mathrm{U}(1)$ or $\mathrm{SU}(3)$ gauge group, which we refer to as dark QED (\texttt{dQED}) and dark QCD (\texttt{dQCD}), respectively. For these models, the coupling strength of the (dark) gauge interaction constitutes an additional parameter. However, as discussed in Sec.~\ref{sec:rescale}, upon neglecting running effects the BSF dynamics factorizes such that the relevant dependence can be expressed in terms of a single parameter which means a one-dimensional grid in $x$ is sufficient. The corresponding physical cross sections for arbitrary values of $m$ and $\alpha$ are then obtained via the rescaling procedure described in Sec.~\ref{sec:rescale}, including an approximate correction to restore running effects. This strategy allows one to efficiently cover a broad class of models with minimal tabulation effort, while retaining full parametric flexibility. The respective models that involve rescaling are annotated with a checkmark in the next-to-last column of Table~\ref{tab:model-coverage}.

\subsection{Numerical implementation}\label{sec:numimpl}

To obtain the tabulated results that \codename uses for interpolation, we numerically implement the expressions introduced in Sec.~\ref{sec:BSFXS} and compute the effective BSF annihilation cross section,
$\langle\sigma v\rangle_{\rm eff,BSF}$, for the models listed in Table~\ref{tab:model-coverage} covering a wide range of temperature parameters, $x\in[10,10^6]$ and (if applicable) masses, $m \in [100, 10^7]\,$GeV\@. Note that in all models using rescaling, the mass is not restricted to the above interval.\footnote{Furthermore, due to the controlled asymptotic behaviour towards large $x$ in the absence of running of $\alpha$, we provide extrapolation of the form $\langle\sigma \vel\rangle_\text{eff,BSF} \simeq a \times x^b$ beyond $x=10^6$. This approximates the contributions of bound states $n>100$, see Sec.~\ref{sec:unitarity}. 
}  
All results are precomputed and stored in tabulated form, due to the complexity and large number of the underlying scattering-to-bound and bound-to-bound transition rates.

For each model, we include bound states with principal quantum numbers up to $n\leq100$ and orbital angular momentum $\ell=0,\dots,n-1$, corresponding to a total of $5050$ bound states with distinct quantum numbers $(n,\ell)$. The degeneracy associated with magnetic quantum numbers is already summed over in our analytical expressions and therefore does not lead to additional distinct contributions. For models in which bound-to-bound transitions are neglected, such as pure (dark) QCD scenarios without additional electroweak interactions, only $\ell=0$ bound states are included.\footnote{
    For bound states with $\ell \geq 1$, decay rates are parametrically suppressed. Specifically, the wave function at the origin vanishes for $\ell>0$ and bound-state decays require higher-order terms in the non-relativistic expansion of the decay matrix element. This leads to a suppression of the decay rate by a factor of order $E_{n\ell}/m$~\cite{Garny:2021qsr}. As a result, the depletion efficiency factors of such states are strongly suppressed and, in particular, dominated by ionisation processes rather than decay. Contributions from $\ell>0$ bound states are therefore systematically subdominant in the scenarios without bound-to-bound transitions and can safely be neglected in our model selection.
}

We verified that inclusion of excitations up to $n=100$ is sufficient to obtain converged results across the tabulated ranges in $m$ and $x$. Specifically, we checked that even for the highest considered $x$ truncating excitations at $n=50$ and $n=100$ results in a percent-level difference in the effective cross section only. We therefore restrict ourselves to $n\leq100$ throughout this work.

For the models involving SM QCD interactions (\texttt{QCD-}\dots), we evaluate the running coupling $\alpha_\text{s}$ using \textsc{RunDec}~3.3~\cite{Herren:2017osy}, including up to five-loop running. The relevant hard, soft and ultrasoft scales entering the BSF, transition and decay rates are chosen as specified in Sec.~\ref{sec:BSFXS}.  
Electromagnetic interactions are treated with a fixed coupling $\alpha_{\rm em}=1/128.9$.

The thermally averaged cross sections are obtained by numerical integration over the relative velocity $v$. We integrate over the range $v_{\rm min}=10^{-5}$ to $v_{\rm max}=10^{0.3}$, where the latter value should be understood as a formal upper bound within the non-relativistic approximation. In practice, contributions from large velocities are exponentially suppressed by the thermal distribution, rendering the precise choice of $v_{\rm max}$ irrelevant for the final result.

All velocity integrals are evaluated using Mathematica with the \texttt{GlobalAdaptive} algorithm, employing \texttt{AccuracyGoal} $=20+2\log_{10}(m)$, \texttt{PrecisionGoal} $=8$,
\texttt{MinRecursion} $=2$ and \texttt{MaxRecursion} $=30$. For highly excited bound states, the integrands exhibit rapid oscillatory behaviour. To ensure numerical stability while maintaining efficiency, the integration is performed in up to $(9+n-\ell)/4$ logarithmically equi-spaced velocity intervals, with low-velocity intervals ignored when they are numerically insignificant (and often unstable or increasingly slow). This choice was found empirically to provide a robust balance between convergence and computational cost.

The tabulated effective cross sections are interpolated using linear interpolation in logarithmic space, where $\langle\sigma v\rangle_{\rm eff,BSF}$ is a smooth and nearly piecewise-linear function of both $m$ and $x$. We provide interpolation routines for all models in Python, C, and Mathematica. It is designed such that future extensions to additional models or denser tables can be accommodated straightforwardly, provided the data are organised in increasing order of $m$ and, for each mass point, increasing $x$. 
With these features, the tool is readily usable for integration into  Boltzmann solvers.\footnote{For instance, within \textsc{micrOMEGAs} our results can be included conveniently via the function \texttt{improveCrossSection}  which allows for temperature-dependent modifications of the annihilation cross section from version~6.3.0 on. Specifically, $\sigma_\text{eff,BSF}(v,T) = \langle \sigma v \rangle_\text{eff,BSF} (m/T)/v$ has to be added to the respective constituents' annihilation cross section.}

\subsection{Limitations from unitarity violation}
\label{sec:unitarity}

Radiative bound state formation is a 2-to-2 scattering process and as such subject to the partial wave unitarity bound
\begin{equation}
\label{eq:unitaritybound}
(\sigma \vel)_\text{uni}^\lp  = \frac{4\pi(\lp+1)}{m^2 \vel} \geq (\sigma v)_\BSF^\lp \equiv \sum_{n,\el} (\sigma \vel)_\BSF^{p,\lp\to n,\el}.
\end{equation} 
As shown previously \cite{Beneke:2024nxh,Binder:2023ckj}, BSF in SU($N$) gauge theories systematically leads to perturbative unitarity violation, even at arbitrarily small coupling strengths, provided sufficiently high $n$ and small $\vel$ are included (see also \cite{Flores:2024sfy,Flores:2025uoh} for recent discussions of unitarisation). Apart from this systematic unitarity violation, our perturbative results naturally become unreliable at large coupling strengths. For \texttt{dQED} models, only this strong-coupling regime can be unreliable, while for \texttt{$($d$)$QCD} one must investigate whether the included velocity region respects the unitarity bound in order to obtain reliable results.

In this section, we provide a detailed analysis of unitarity bounds in all parameter regions accessible in \codename and deduce appropriate conservative conditions where warnings are produced.
We restrict ourselves to $s$- and $p$-wave bounds since the former has been found to be systematically strongest at low velocities \cite{Binder:2023ckj} while the latter applies to models without bound-to-bound transitions (\texttt{QCD-S}, \texttt{QCD-F}, \texttt{dQED-SnoTr}, \texttt{dQED-FnoTr}) where only capture into $\lp=1\to\el=0$ is relevant.

\begin{figure}[t]
\begin{center}
\includegraphics[width=0.8\textwidth]{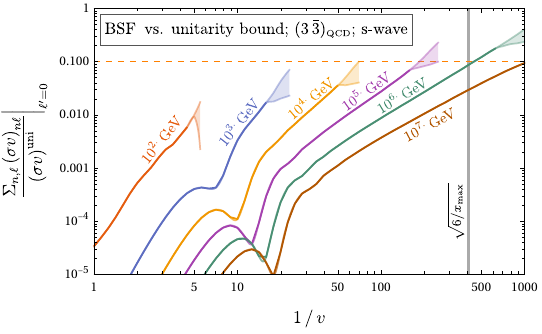}
\caption{\label{fig:UViQCD}
Ratio of the summed $s$-wave BSF cross section to the corresponding unitarity bound,
$(\sigma v)_\BSF^{\ell'=0}/(\sigma v)_\text{uni}^{\ell'=0}$, for SM QCD, plotted as a function of inverse velocity.
The vertical gray line indicates the virial velocity corresponding to the highest $x$ included in our tables,
$v=\sqrt{6/x_\text{max}}$.
The coloured region, where the curves begin to fan up at low velocities, indicates the onset of deviations between two prescriptions for the running coupling at energy scales below 1\,GeV
(plateau versus cutoff).
}
\end{center}
\end{figure}

For \texttt{QCD} models, the running of the coupling is fixed, $\alpha(\mu)=\alpha_\text{QCD}(\mu)$, and we show the inclusive cross-section divided by the unitarity bound, $(\sigma v)_\BSF^{\lp=0}/ (\sigma \vel)_\text{uni}^{\lp=0}$, in \fig~\ref{fig:UViQCD} for various mass values from our data set, including the minimal and maximal masses ($m=10^2$\,GeV and $10^7$\,GeV) plotted over inverse relative velocity. We overlay the two lines of ``cutoff'' and ``plateau'' treatments of large couplings for each mass. 
This is a simple but effective indicator to highlight strong coupling regions near the QCD Landau pole where perturbative calculations are not reliable. We only show the onset of the large coupling regime for each mass for clarity of the figure.
The strict upper bound from unitarity is found at $1$ and we also highlight the level of $10\%$ of the unitarity bound by a horizontal dashed line. The vertical gray line indicates the virial velocity of the largest $x=x_\text{max}=10^6$ included in our tabulated data set. Due to the combination of Boltzmann suppression and non-relativistic repulsion in the initial colour-octet state, one expects only a narrow velocity range around the virial velocity to contribute to the thermal average at low $v$ (i.e.~high $n$). Clearly, all results remain well below the unitarity bound within our analysis when avoiding also the strong coupling regime. Hence, we do not expect corrections from unitarisation beyond the accuracy of our perturbative calculation \cite{Flores:2024sfy}.

\begin{figure}[t]
\begin{center}
\includegraphics[width=0.8\textwidth]{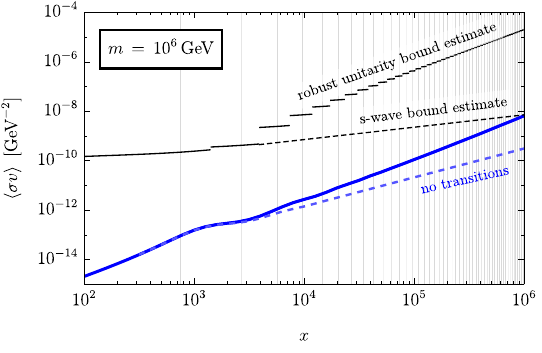}
\caption{\label{fig:approximaterobust}Comparison of the effective thermally averaged cross section $\langle \sigma v \rangle_\text{eff,BSF}$ to estimates of the corresponding unitarity bounds. The blue solid and blue dashed curve correspond to the models with (\texttt{QCD-SU}) and  without (\texttt{QCD-S}) bound-to-bound transitions, respectively, choosing $m=10^6\,$GeV. The black solid line corresponds to the most robust unitarity bound summing over all potentially relevant partial wave contributions at a given $x$ (see text for details) while the black dashed line displays the thermally averaged $s$-wave contribution.}
\end{center}
\end{figure}

Fig.~4 illustrates these aspects directly at the level of the thermally averaged effective cross section, $\langle\sigma v\rangle_{\text{eff,BSF}}$, as a function of the temperature parameter $x$, which is the quantity tabulated and provided by \codename. In the absence of bound-to-bound transitions (dashed blue curve), annihilation proceeds through capture into the decaying $\ell=0$ bound states from $\ell'=1$ scattering states, such that the effective cross section is bounded by the corresponding $p$-wave unitarity limit, 
$
\langle\sigma v\rangle_{\text{eff,BSF}}^{\text{no-trans}}
\le \langle(\sigma v)_{\text{BSF}}^{\ell'=1}\rangle
\le \langle(\sigma v)_{\text{uni}}^{\ell'=1}\rangle
\propto x^{1/2},
$
where the first inequality follows from the fact that $0 \le R_i \le 1$ and
we used $f_B(\omega)=0$ for the thermal averaging of $(\sigma v)_{\text{uni}}$, which provides the most aggressive choice as it neglects all Bose enhancement from \eq~(\ref{eq:defthermalaverage}) in the thermal average of the left-hand side of \eq~(\ref{eq:unitaritybound}). This comes in addition to showing the $s$-wave bound which itself is a factor of 1/3 more stringent than the $p$-wave bound which would be applicable to the no transitions result. While $\langle\sigma v\rangle_{\text{eff,BSF}}^{\text{no-trans}}$ grows considerably more steeply with $x$ than the unitarity bound, it remains far below the bound throughout the range covered by our tabulation.

When bound-to-bound transitions are included (blue solid curve), $\langle\sigma v \rangle_{\text{eff,BSF}}$ grows even more rapidly with $x$. In this case, as an increasing number of excited states and higher angular momenta become relevant at large $x$, a growing number of partial waves must be summed to obtain applicable unitarity bounds. Hence, this resulting unitarity constraint relaxes substantially relative to the $s$-wave bound. To illustrate this behaviour, we display an estimate of the most robust unitarity bound obtained by summing all partial-wave unitarity limits $(\sigma \vel)_\text{uni}^\lp$ up to the parametrically estimated highest contributing $\lp$. 
Specifically, we compute $ \sum_{\ell'=0}^{\ell'_{\max}} \langle(\sigma v)_{\text{uni}}^{\ell'}\rangle|_{f(\omega_{n_{\max}})},
$ where $\lp_{\max}$ is the  $\lp_{\max}=\el_{\max}-1=n_{\max}-2$ for a given $x$, with $n_{\max}$ implicitly defined through $ \frac32 T = | {E_{n_{\max}(x)}} | $.
We perform the thermal average with $\omega_{n_{\max}}=m \vel^2/4(1+\zeta_n^2)$ to arrive at the most robust bounds which we show as the solid black curve in the figure where each step is caused by inclusion of the next highest $\lp$. We always include at least the $s$-wave bound. The effective cross section remains far below this bound (and in fact even below the $s$-wave unitarity limit) in the entire tabulated range.

\begin{figure}[t!]
\begin{center}
\includegraphics[width=0.5\textwidth]{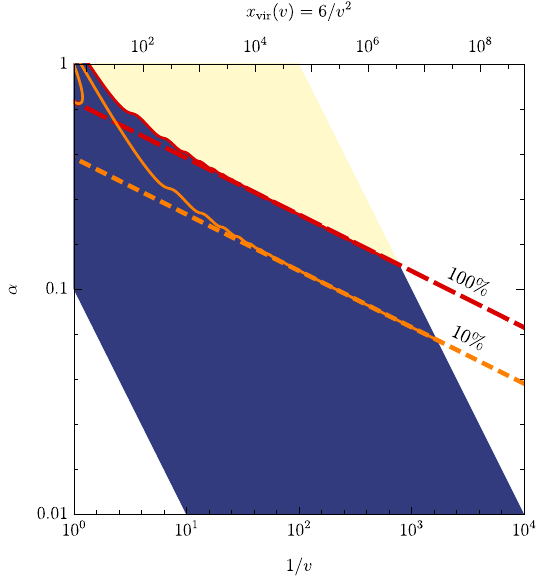}
\caption{\label{fig:UVidarkSU3}
Unitarity violation in dark QCD with constant coupling.
Problematic and unproblematic regions are shown in yellow and blue, respectively, including bound states up to $n=4000$.
The red and orange contours indicate where $100\%$ and $10\%$ of the $s$-wave unitarity bound $(\sigma v)^{\ell=0}_\text{uni}$ are reached.
Dashed lines show an extrapolation of the low-$\vel$ boundary of the unitarity-violating region to $n_\text{max}\to\infty$ based on the displayed data, see text for details. 
}
\end{center}
\end{figure}

For \texttt{dQCD} models of constant couplings, our analysis of possible unitarity violation  becomes more involved as it must be performed for general values of $\alpha$ and $\vel$. However, exploiting the rescaling relations described in Sec.~\ref{sec:rescale} significantly simplifies this task. The resulting two-dimensional parameter plane is shown in \fig~\ref{fig:UVidarkSU3}, where contours corresponding to $100\%$ (red solid curve) and $10\%$ (orange solid curve) of the $s$-wave unitarity bound are indicated.  Accordingly, the yellow region marks parameter points where $s$-wave unitarity is violated by perturbative BSF. 
The entire coloured regions in the 2-dimensional plane is derived from a tabulation of only a single $\alpha$ and exploiting the rescaling feature, see Section~\ref{sec:rescale}, which explains the rhombic shape of the displayed data set.
No explicit data was computed in the white regions, however, the functional behaviour of $(\sigma \vel)_\BSF^\ell$ is sufficiently well understood to allow for a reliable extrapolation. In particular, the summation over principal quantum numbers converges to a smooth behaviour at low velocities~\cite{Garny:2021qsr,Beneke:2024nxh,Lederer:2024lkl} and can be well approximated by a simple monomial in this regime. Too low $\vel$ must be excluded as $n>4000$ would become important. We perform this extrapolation using a linear fit in double-logarithmic space and show the corresponding extrapolations to lower $v$ as dashed lines for the $100\%$ and $10\%$ contours. Specifically, we find $\log_{10}(\alpha)= -0.166-0.251 \log_{10}(1/\vel)-0.250 \log_{10} (\rho)$, where $\rho$ denotes the desired ratio of the unitarity bound: 10\% (orange dashed) and 100\% (red dashed). 

\codename raises warnings based on this fit, whenever the 10\% approximation of the unitarity bounds is exceeded in \texttt{dQCD} models. This together with the above analysis of QCD BSF, \codename can be used reliably without concern of large corrections from unitarity violation.

\section{Phenomenological application}
\label{sec:application}

To illustrate a typical use case of \codename in a cosmological setting, we  revisit the superWIMP scenario studied in Ref.~\cite{Binder:2023ckj}. We consider a singlet Majorana fermion $\chi$ as the dark matter candidate and a coloured scalar mediator $\tilde q$ with the gauge quantum numbers of a right-handed quark. The latter is subject to BSF\@. The interaction between the dark sector and the SM is described by the Yukawa-type interaction
\begin{equation}
  {\cal L}_{\text{int}} = \lambda_\chi\, \tilde q\, \bar q_R\, \chi + \text{h.c.}\,,
\end{equation}
while the mediator $\tilde q$ couples to gluons and hypercharge gauge bosons through the usual kinetic term. We assume $m_{\tilde q}>m_\chi$, such that $\tilde q$ is unstable and eventually decays into the dark matter particle. Depending on the size of the decay (or conversion) rate $\Gamma_{\tilde q\to q\chi}$, the dark matter abundance can be set by coannihilation, conversion-driven freeze-out or the superWIMP mechanism~\cite{Garny:2017rxs,DAgnolo:2017dbv,Covi:1999ty,Feng:2003uy,Garny:2018ali}. In the latter case, which we focus on here, the mediator undergoes its own freeze-out and the remaining population subsequently decays out of equilibrium into dark matter.

As shown in Ref.~\cite{Binder:2023ckj}, the presence of Coulombic bound states of the coloured and electrically charged mediator can drastically change this picture. Bound-state formation, transitions among excited levels and bound-state decays lead to a super-critical effective annihilation cross section for the mediator, such that its abundance does not truly freeze out but continues to decrease until the decay $\tilde q\to q\chi$ becomes efficient. The final dark matter density then depends on the mediator lifetime and hence on $\Gamma_{\tilde q\to q\chi}$, contrary to the standard superWIMP paradigm where a constant abundance after freeze-out is assumed.

In the present work, we utilize \codename to compute the effective thermally averaged cross section\footnote{Specifically, we compute $\langle \sigma v \rangle_\text{eff} = \langle (\sigma v)_\text{ann} \times S_0  \rangle + \langle \sigma v \rangle_\text{eff,BSF}$ using the known analytic expressions for the $s$-wave mediator-pair annihilation cross section, $ (\sigma v)_\text{ann} $, and Sommerfeld enhancement factor, $S_0$, see e.g.~Appendix B of Ref.~\cite{Binder:2023ckj}.} entering the Boltzmann equation for the mediator abundance and hence rapidly obtain predictions for the relic density. Concretely, we
consider two benchmark choices for the mediator hypercharge: a bottom-philic case with $Q=-1/3$ and a top-philic case with $Q=2/3$. (In Ref.~\cite{Binder:2023ckj} the former was considered only.) In both cases the long-range potential and bound-state formation are dominated by QCD, while the different electric charges control the strength of electromagnetic dipole transitions among excited levels. The larger charge in the top-philic setup leads to enhanced bound-to-bound transition rates and thus to a more efficient depletion of the mediator abundance at late times. 

\begin{figure}[t]
\begin{center}
\includegraphics[width=0.49\textwidth]{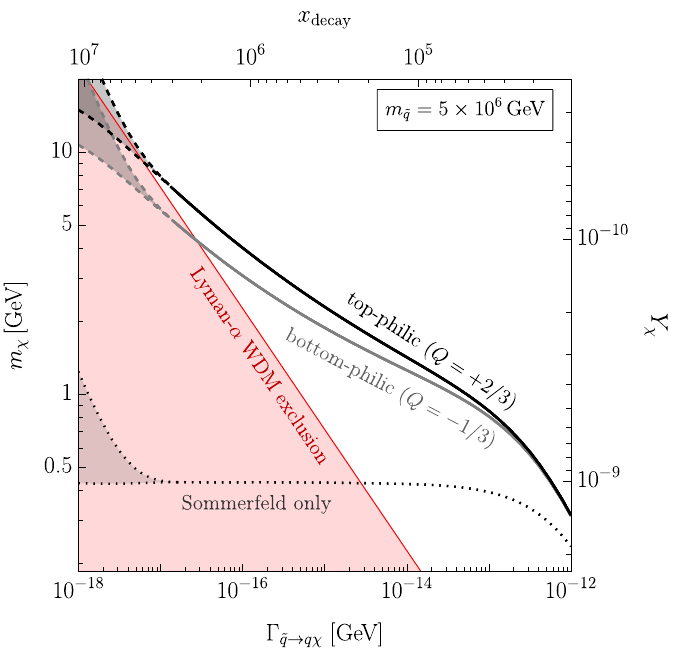}
\includegraphics[width=0.49\textwidth]{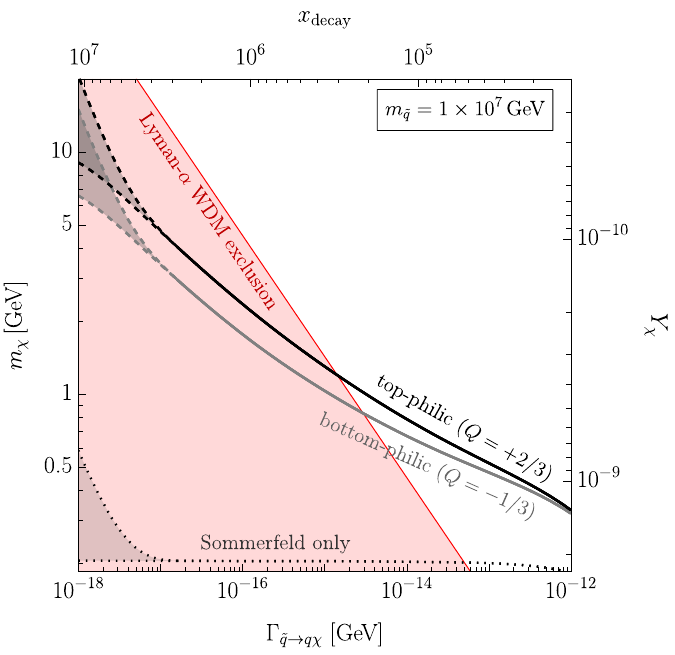}

    \caption{\label{fig:Gamma_mchi} Relic density contours in parameter space of decay width and dark matter mass for $m_{\tilde q}=5\times 10^6$\,GeV (left panel) and $m_{\tilde q}=10^7$\,GeV (right panel). We show results for the bottom-philic ($Q=-1/3$, gray) 
    and top-philic model ($Q=+2/3$, black) as well as without bound state effects (considering leading-order mediator pair-annihilation with Sommerfeld enhancement only) for which the two model predictions are equal. The lines fanning up at low decay rates bracket possible values of $Y_\chi^0$ by assuming $Y_\chi^0 = Y_\chi(T=1\,\text{GeV})$ (upper dashed boundary) and $Y_\chi^0 = Y_\chi(1\,\text{GeV}) + Y_{\tilde q}(1\,\text{GeV})$ (lower dashed boundary), thereby providing a conservative estimate of the uncertainty associated with late decays at $T \lesssim 1\,\text{GeV}$.}
\end{center}
\end{figure}

Figure~\ref{fig:Gamma_mchi} shows the result for two different mediator masses $m_{\tilde q}=5\times 10^6$\,GeV (left panel) and $10^7$\,GeV (right panel). For each value of the decay rate $\Gamma_{\tilde q\to q\chi}$ the plots show the corresponding dark matter mass $m_\chi$ that reproduces the observed relic density $\Omega_\chi h^2 \simeq 0.12$ in the bottom- and top-philic scenarios. The qualitative behaviour is the same as in Ref.~\cite{Binder:2023ckj}: in the superWIMP regime the mediator abundance continues to decrease until $H \sim \Gamma_{\tilde q\to q\chi}$, so that a larger mediator lifetime (smaller $\Gamma_{\tilde q\to q\chi}$) leaves fewer mediators to decay and hence requires a larger $m_\chi$ to match the observed density. Compared to the bottom-philic case, the more efficient depletion in the top-philic case implies a systematically larger $m_\chi$ for the same decay rate.

For reference, we also show the result including perturbative pair annihilation  with Sommerfeld enhancement only. In this case, the mediator abundance -- and therewith the $m_\chi$ that matches the measured relic density -- stays constant over a large range of $\Gamma_{\tilde q\to q\chi}$ in which the mediator freezes out and decays well after. This result underpins the importance of (excited) bound states causing an order-of-magnitude effect on $m_\chi$ for late mediator decays which, however, still take place at temperatures above 1\,GeV, i.e.~in the perturbative regime of QCD\@.

Finally, in Fig.~\ref{fig:Gamma_mchi} we also overlay the constraints from structure formation. The delayed and boosted production of dark matter induces a hardening of its momentum distribution, i.e.~warm dark matter, which can be constrained by Lyman-$\alpha$ forest observations. Utilizing the results of Ref.~\cite{Decant:2021mhj}, for the mediator mass $m_{\tilde q}=5\times 10^6\,\text{GeV}$, we find that part of the bottom-philic curve at small decay rates is excluded by the Lyman-$\alpha$ bound, while the entire top-philic curve remains allowed in the perturbative QCD regime (solid part of the two curves). This illustrates the phenomenological relevance of electromagnetic transitions which constitute the only significant difference between the two models in this mass regime. For larger $m_{\tilde q}$ the Lyman-$\alpha$ bound swiftly approaches shorter lifetimes and start constraining significant parts of the parameter space. Toward smaller masses, the bound shifts to the left quickly losing sensitivity to the superWIMP scenario within the perturbative QCD regime.  

Importantly, obtaining these curves with \codename only requires interpolating tabulated effective cross sections as input for the Boltzmann equations making such phenomenological explorations feasible and many orders of magnitude faster than the original computation of Ref.~\cite{Binder:2023ckj}.

%
\section{Conclusion}
\label{sec:concl}
In this work, we presented \codename, a fast and lightweight numerical interpolation tool for computing bound-state effects on annihilation processes of new physics particles in the early Universe. While our primary motivation is dark matter phenomenology, the framework  applies more generally to new particles interacting via long-range forces. The code is based on precomputed tables of effective thermally averaged annihilation cross sections that consistently include bound-state formation, dissociation, bound-to-bound transitions and decay. Highly excited bound states up to principal quantum number $n=100$ are taken into account, allowing for a reliable description of late-time annihilation at large values of the temperature parameter $x=m/T$.

A key feature of \codename is the exploitation of rescaling properties of the effective cross section in scenarios where the long-range force coupling can be treated as approximately constant. This allows a wide class of models with unbroken $\text{U}(1)$ or $\text{SU}(3)$ gauge interactions to be covered efficiently with minimal tabulation effort. For models involving SM QCD, where running effects are significant, \codename instead relies on direct tabulation and two-dimensional interpolation in mass and temperature, without invoking rescaling. In both cases, the tool provides fast and reliable access to bound-state enhanced annihilation rates over broad ranges of masses and temperatures, making it well suited for repeated use in Boltzmann solvers and parameter scans.

The large contribution of highly excited bound states at late times naturally raises concerns about perturbative unitarity. We addressed this issue in detail. For particles charged under SM QCD, we showed that the summed $s$-wave BSF cross section remains safely below the corresponding unitarity bound -- by at least an order of magnitude -- when restricting to perturbative couplings $\alpha_\text{s} \le \alpha_\text{s}(\mu\!=\!1\,\mathrm{GeV})$ and within the tabulated range $x \le 10^6$. We further demonstrated that the thermally averaged effective cross section remains below the thermally averaged unitarity bound in this regime, even when adopting conservative estimates based on the $s$-wave contribution only. We therefore do not expect any phenomenologically relevant impact from unitarization effects within the domain of applicability of \codename.  

For dark QCD scenarios with constant couplings, the broader parameter coverage includes regions where unitarity bounds can be approached; these cases are explicitly flagged by \codename whenever the effective cross section reaches within 10\% of the corresponding bound.

As an illustration, we applied \codename to two superWIMP scenarios with a coloured bottom- and top-philic mediator. We demonstrated that bound-state effects can qualitatively alter the late-time annihilation dynamics and the resulting dark matter abundance, with important implications for structure-formation constraints from Lyman-$\alpha$ observations. In particular, we found that the larger rate for transitions among excited levels pushes the top-philic model further away from the Lyman-$\alpha$ bounds. 

The tool is publicly available on \href{https://github.com/bsffast/BSFfast}{GitHub}\footnote{\href{https://github.com/bsffast/BSFfast}{\url{https://github.com/bsffast/BSFfast}}} and provides interfaces in C, Python and Mathematica, allowing for its integration into Boltzmann solvers such as \textsc{micrOMEGAs} and \textsc{MadDM}. Future extensions of \codename may include additional gauge groups or further classes of long-range interactions. We expect \codename to be a useful resource for a broad range of early-Universe studies involving bound-state dynamics within dark matter models and beyond.

\section*{Acknowledgements}

We thank Andreas Goudelis and Alexander Pukhov for helpful exchanges on the use of \codename within \textsc{micrOMEGAs} and Martin Napetschnig for valuable discussions. TB~acknowledges support from the Advanced ERC grant ERC-2023-ADG-Project EFT-XYZ (Prof.~Nora Brambilla).
SL is supported by the National Science Centre (Poland) under the research Grant No.~2021/42/E/ST2/00009. SL~gratefully acknowledges the continued access to compute resources of TUM during the completion of this work.

\appendix
\section{Code usage}
\label{app:usage}

\textsc{BSFfast} is publicly available via the \href{https://github.com/bsffast/BSFfast}{GitHub repository}.
The code is distributed with Python, C, and Mathematica interfaces.
For definiteness, we describe the Python interface below; the other
implementations provide analogous functionality.

\subsection{Public interface}

The Python implementation provides a single public function
\begin{equation}
\texttt{fastXS(model, x, m, alpha=None)}\,,\nonumber
\end{equation}
which returns the effective thermally averaged BSF cross section $\langle \sigma v \rangle_{\mathrm{eff,BSF}} $ in units of GeV$^{-2}$.
The arguments are:
\begin{itemize}
\item \texttt{model}: string specifying the model (see Table~\ref{tab:model-coverage}),
\item $\texttt{x} = m/T$: temperature variable,
\item \texttt{m}: particle mass in GeV,
\item \texttt{alpha}: coupling specification or scheme selector.
\end{itemize}
The argument \texttt{alpha} can take several forms:
\begin{itemize}
\item \texttt{None}: use the default coupling or scheme,
\item a float: constant coupling $\alpha$,
\item a callable \texttt{alpha(q)}: running coupling,
\item a string \texttt{"cutoff"} or \texttt{"plateau"}: infrared scheme selector (SM QCD models).
\end{itemize}
\medskip
A minimal example reads:
\smallskip
\begin{verbatim}
from BSFfast import fastXS

x = 200.0
m = 1000.0  # GeV

xs = fastXS("dQCD-S", x, m, alpha=0.1)
\end{verbatim}
\smallskip

\subsection{Two classes of models}

The implemented models fall into two structurally distinct classes.

\paragraph{(i) Two-dimensional grid models.}
For SM QCD models (e.g.\ \texttt{QCD-SU}), the cross section is tabulated
on a two-dimensional grid in $(m,x)$ using the running SM coupling.
Interpolation is performed in $\log m$ and $\log x$.

Two infrared prescriptions are provided and can be selected via the
\texttt{alpha} argument:
\begin{equation}
\texttt{alpha = "cutoff"} \quad (\text{default}), 
\qquad
\texttt{alpha = "plateau"}.\nonumber
\end{equation}
If \texttt{alpha=None}, the default \texttt{"cutoff"} scheme is used.

\paragraph{(ii) Rescaled models.}
For dark-sector QCD and QED models (e.g.\ \texttt{dQCD-S}, \texttt{dQED-F}),
the tables are generated at a fixed reference mass $m_0$ and coupling
$\alpha_0$. At runtime, results are obtained via the rescaling relations
derived in Sec.~\ref{sec:rescale}. The coupling can be specified explicitly, e.g.
\begin{verbatim}
fastXS("dQCD-S", x, m, alpha=0.15)
\end{verbatim}
\medskip
Alternatively, approximate running-coupling effects may be included
by providing a callable:
\begin{verbatim}
def alpha_running(q):
    return ...

fastXS("dQCD-S", x, m, alpha_running)
\end{verbatim}
\medskip
In this case the coupling is evaluated at the characteristic scale
\begin{equation}
q = m \sqrt{C/x},
\end{equation}
as discussed in Sec.~\ref{sec:rescale}. 
In the public Python implementation, the default choice is
$C=2$ (see the parameter \texttt{C\_POT\_SCALE} in \texttt{BSFfast.py}).
This prescription provides an efficient approximation of moderate
running effects without tabulating additional dimensions.

\subsection{QED models}

The QED models require a small amount of additional clarification.

\paragraph{QED-S.}
The model \texttt{QED-S} is implemented as a convenience alias of
\texttt{dQED-S}. By default, the electromagnetic coupling
$\alpha_{\rm em}=1/128.9$ is used. The user may override the coupling,
in which case the model becomes equivalent to \texttt{dQED-S} with the
specified $\alpha$.

\paragraph{QED-F.}
In contrast, \texttt{QED-F} is not simply \texttt{dQED-F} evaluated at
$\alpha_{\rm em}$. For fermionic constituents with an abelian gauge
interaction, bound-state decay into SM fermion pairs via an off-shell
photon contributes significantly to the effective rate.

To account for the opening of the $t\bar t$ channel, two separate tables
are provided:\linebreak
\texttt{QED-FexclTop}~and \texttt{QED-FinclTop}.
The default model \texttt{QED-F} result is obtained via a phase-space-weighted
interpolation between the two,
\begin{equation}
\langle\sigma v\rangle_{\texttt{QED-F}}
=
(1-w)\,\langle\sigma v\rangle_{\texttt{exclTop}}
+
w\,\langle\sigma v\rangle_{\texttt{inclTop}},
\end{equation}
where $w(m)$ approximates the top-pair phase-space factor.

This prescription provides a good approximation away from the threshold
region. Threshold resummation effects are not included and may modify
the decay rate near $m \sim m_t$. For full control, the user may call
the models \texttt{QED-FexclTop} and \texttt{QED-FinclTop} explicitly.

\subsection{Warnings and validity}

For rescaled dark QCD models, the code monitors the proximity to the
unitarity bound discussed in Sec.~\ref{sec:unitarity}. 
Based on the fit described there, a warning is issued if the estimated
cross section exceeds
\begin{itemize}
\item $10\%$ of the unitarity limit,
\item $100\%$ of the unitarity limit.
\end{itemize}
Each warning is emitted at most once per model during runtime.

For the model \texttt{QED-F}, an informational warning is displayed
to indicate that the default result is obtained from a phase-space-weighted
interpolation between the \texttt{exclTop} and \texttt{inclTop} tables.

Interpolation is performed in log–log space.
Outside the tabulated range, extrapolation is used.
Users should ensure that parameter choices remain within the regime
of validity discussed in the main text.

\providecommand{\href}[2]{#2}\begingroup\raggedright\endgroup

\end{document}